\documentstyle[epsfig]{basi}
\begin{document}
\title[$UBVRI$ CCD photometry of the OB associations Bochum 1 and Bochum 6]{$UBVRI$ CCD photometry of the OB associations Bochum 1 and Bochum 6}

\author[R. K. S. Yadav and Ram Sagar]%
       {R. K. S. Yadav\thanks{E-mail: rkant@upso.ernet.in} and Ram Sagar\thanks{E-mail: sagar@upso.ernet.in} \\  
        State Observatory, Manora Peak, Naini Tal 263 129}
\pubyear{2003}
\setcounter{page}{1}
\maketitle
\label{firstpage}
\begin{abstract}
We report the first deep $UBVRI$ CCD photometry of 2460 stars in the field of two poorly 
studied OB associations Bochum 1 and Bochum 6. We selected 15 and 14 probable members in 
Bochum 1 and Bochum 6 respectively using photometric criteria and proper motion data of 
Tycho 2. Our analysis indicates variable reddening having mean value of $E(B-V)=$ 0.47$\pm$0.10 
and 0.71$\pm$0.13 mag for Bochum 1 and Bochum 6 respectively. Using the zero-age main-sequence 
fitting method, we derive a distance of 
2.8$\pm$0.4 and 2.5$\pm$0.4 Kpc for Bochum 1 and Bochum 6 respectively. We obtain an age 
of 10$\pm$5 Myrs for both the associations from isochrone fitting. In both associations 
high and low mass stars have probably formed together. Within the observational uncertainties, 
mass spectrum of the both associations appears to be similar to the Salpeter's one.  
\end{abstract}

\section{Introduction}
\label{sec:intro}
The study of young stellar systems in a galaxy provides useful information on its 
structure and formation history. Most of the stars are born as members of 
stellar groups of different types (Gomez et al. 1993; Massey et al. 1995). 
Stellar associations being typically loose 
stellar systems (Blaauw 1964) characterized by their bright blue stellar populations, are 
considered tracers of the distribution of the youngest population in a galaxy.
Moffat and Vogt (1975) suggest that Bochum 1 is a group of 9 OB stars in which 
8 stars show a sequence in the $UBV$ colour-colour (CC) and colour-magnitude (CM) diagram 
and hence can be considered at the same distance of 4.06 Kpc from Sun. They derived a 
mean reddening of $E(B-V)=$ 0.45$\pm$0.06 mag in the direction of the object. Recently 
Fitzsimmons (1993) presented the CCD Str\"omgren uvby photometry of 9 stars in the 
Bochum 1 region and found that Bochum 1 is 2$\times$10$^{7}$ yrs old and 
located at a slightly larger distance of 4.8 Kpc. Moffat and Vogt (1975) also studied Bochum 6 
for the first time and conclude that this is a group of 5 OB stars. They also
indicated that the object appears to coincide with the HII region S 309 for
 which Georgelin et al. (1973) gives photometric distance of 6.30 Kpc and
Kinematic distance of 2.24$\pm$0.36 Kpc. Moffat \& Vogt (1975) derived a mean value of
reddening $E(B-V)$ = 0.70$\pm$0.10 mag, and distance 4.0 Kpc which lies almost
midway between Georgelin et al.'s two independent estimates. Other basic informations 
about both Bochum 1 and 6 are given in Table 1. For both the objects available photometric 
studies are limited to the stars brighter than $V\sim$ 12.5 mag. We therefore
obtained first deep $UBVRI$ CCD photometric data and used them in combination with 
the available kinematical data for better estimation of basic parameters, such 
as reddening, distance and age. Such parameters are valuable for understanding the 
disk sub-system which these associations belong to. Although, deep photometric data is not 
much useful in our study but it is valuable in modelling of the Galaxy. The 
next section presents the observations and data reduction while the membership, 
reddening, distance and age etc. of both the objects are determined in the 
remaining part of the paper.

\begin{table}
\caption{General information of the objects under study, taken from Dias et al. (2002).}
\label{table1}
\vspace{0.5cm}
\footnotesize
\begin{center}
\begin{tabular}{|c|cccccc|}
\hline
\hline
Objects &$\alpha_{2000}$&$\delta_{2000}$&Radius&Distance&$E(B-V)$&log(age)\\
&&&(arc min)&(Kpc)&(mag)&\\ \hline
Bochum 1&$06^{h}25^{m} 27^{s}$&$19^{d} 46^{\prime} 15^{\prime\prime}$&-&2.8&0.50&6.7\\
Bochum 6&$07^{h}32^{m}00^{s}$&$-19^{d} 26^{\prime} 27^{\prime\prime}$&5.0&3.9&0.70&7.0\\
\hline
\end{tabular}
\end{center}
\end{table}

\begin{figure}
\begin{center}
\hbox{
}
\caption {The $13^{\prime}.0\times13^{\prime}.0$ area of Bochum 1 and Bochum 6 taken from DSS. 
North is up and east is left.}
\label{fig1}
\end{center}
\end{figure}

\section{$UBVRI$ CCD Observations and data reductions }
\label{sec:using}

The CCD broad band photometric $UBV$ (Johnson), $RI$ (Cousins) observations were carried 
out using 2K$\times$2K CCD system at the f/13 Cassegrain focus of the Sumpurnanand 104-cm 
telescope of the State observatory, Nainital. The read-out noise and gain of the CCD 
system are 5.3 e$^{-}$ and 10 e$^{-}$/ADU respectively. The 0.$^{\prime\prime}$36/pixel 
plate scale resulted in a field of view of 12$^{\prime}$.3$\times$12$^{\prime}.3$. 
 For the accurate photometric measurements of fainter stars, 2 to 3 deep exposures were 
taken in each passband in 2$\times$2 pixel binning mode. Further details of the observations 
are listed in Table 2. The covered region is shown in Fig 1 where DSS images is presented 
for both Bochum 1 and Bochum 6. 

The data have been reduced by using the IRAF, MIDAS and DAOPHOT software packages. The 
instrumental signatures were removed using bias and flats taken during the observing run 
by means of standard IRAF routines. Further reductions including profile magnitudes of 
the stars were performed using the DAOPHOT (Stetson 1987) in the MIDAS environment. To 
construct a point spread function (PSF) for the entire CCD frame on each exposure, we 
used typically 50 well isolated stars. 

We observed SA 98 (Landolt 1992) standard field in $UBVRI$ for calibrating the observations 
of Bochum 1 and 6. We used 7 standard stars for calibration purpose which are having 11$-$15 
mag range in $V$ and 
0.1$-$2.0 mag range in $(V-I)$ colour. Thus the stars used for calibration cover a wide range 
in brightness as well as in colour. This standard field is also observed at different 
airmasses to obtain a reliable estimate of nightly atmospheric extinction coefficients.

The calibration equations obtained by observing standard field are :

\begin{center}
$u$ $=$ $U + 6.66\pm0.01 - (0.02\pm0.01)(U-B) + 0.45X$\\
$b$ $=$ $B + 4.62\pm0.01 - (0.04\pm0.01)(B-V) + 0.23X$\\
$v$ $=$ $V + 4.22\pm0.01 - (0.03\pm0.01)(B-V) + 0.13X$\\
$r$ $=$ $R + 4.15\pm0.01 - (0.04\pm0.01)(V-R) + 0.09X$\\
$i$ $=$ $I + 4.67\pm0.01 - (0.12\pm0.01)(R-I) + 0.05X$\\
\end{center}

\begin{table}
\caption{Journal of $UBVRI$ photometric observations of Bochum1 and Bochum 6. N is the 
number of stars measured in the filter.}
\label{table2}
\vspace{0.5cm}
\footnotesize
\begin{center}
\begin{tabular}{ccccc}
\hline
\hline
Objects&Date&Exp. Time&Filters&N\\ \hline
Bochum 1&16 Jan 01&1800$\times$3, 300&$U$&750\\
$\alpha_{2000}=06^{h}25^{m}27^{s}$&,,&1200$\times$3, 240&$B$&993\\
$\delta_{2000}=+19^{d}46^{\prime}15^{\prime\prime}$&,,&900$\times$3, 240&$V$&997\\
&''&900$\times$3, 240&$R$&1000\\
&''&300$\times$3, 60$\times$2&$I$&1000\\
Bochum 6&03 Jan 00&1500$\times$3, 900&$U$&270\\
$\alpha_{2000}=07^{h}32^{m}00^{s}$&''&900$\times$3, 180&$B$&1400\\
$\delta_{2000}=-19^{d}26^{\prime}27^{\prime\prime}$&''&600$\times$3, 120&$V$&1430\\
&''&240$\times$3, 60&$R$&1430\\
&''&240$\times$3, 60&$I$&1460\\
\hline
\end{tabular}
\end{center}
\end{table}
where $U, B, V, R$ and $I$ are standard magnitudes, $u, b, v, r$ and $i$ are the 
instrumental one corresponding to exposure time of 1 sec, and $X$ is the airmass.
The plots of colour equations are shown in Fig 2. The errors determined by least 
square fitting in zero points and colour coefficients are $\sim$ 0.01. In Table 3, we 
list the internal errors as a function of brightness estimated on the S/N ratio 
of the stars as outputed by ALLSTAR programme of the DAOPHOT. There are 1000 and 1460 
stars measured in the region of Bochum 1 and Bochum 6 respectively. The (X, Y) pixel 
coordinates and $UBVRI$ magnitudes of the stars are available in electronic form from 
authors as well as from WEBDA site{\footnote{http://obswww.unige.ch/webda/}}. 

\begin{figure}
\begin{center}
\hspace{-0.0cm}\psfig{figure=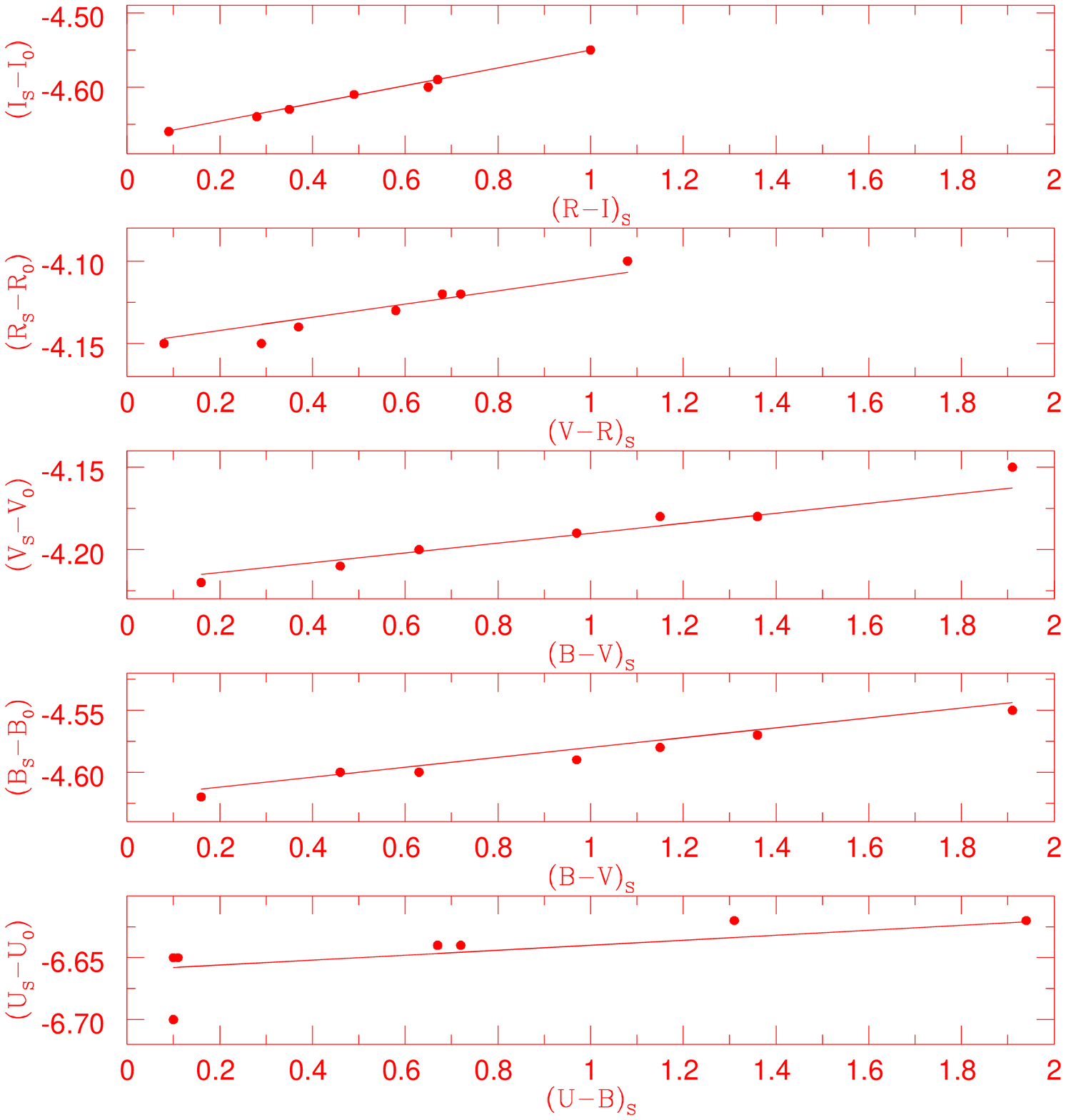, height=8.5cm, width=6.5cm}
\caption {Plots of colour equations. Least square fitting is presented by the solid line.}
\end{center}
\end{figure}

\begin{table}
\centering
\caption{Internal photometric errors as a function of brightness. $\sigma$ is
the standard deviation per observation in magnitude.}
\vspace{0.5cm}
\begin{tabular}{cccccc}
\hline\hline
Magnitude range& $\sigma$$_{U}$&$\sigma$$_{B}$&$\sigma$$_{V}$  &$\sigma$$_{R}$ &
$\sigma$$_{I}$\\
\hline
$\le$13.0&0.011&0.002&0.005&0.003&0.009\\
13.0 - 14.0&0.019&0.006&0.010&0.008&0.019\\
14.0 - 15.0&0.021&0.007&0.012&0.014&0.014\\
15.0 - 16.0&0.023&0.012&0.014&0.017&0.014\\
16.0 - 17.0&0.025&0.012&0.016&0.017&0.017\\
17.0 - 18.0&0.026&0.014&0.016&0.018&0.019\\
18.0 - 19.0&0.026&0.019&0.018&0.018&0.021\\
19.0 - 20.0&0.028&0.025&0.022&0.024&0.023\\
\hline
\end{tabular}
\end{table}

\section{The present study}
\subsection{Apparent colour-magnitude diagrams}
 
The $V$ vs $(U-B)$, $(B-V)$, $(V-R)$ and $(V-I)$ CM diagrams for all the 
stars detected in the region of Bochum 1 and Bochum 6 are shown in Fig 3. In $V$, $(U-B)$ and $V$, 
$(B-V)$ diagrams we have also plotted the brighter stars observed by Moffat \& Vogt (1975)
 but not by us. In the CM diagram of Bochum 1, there is a poorly populated
MS from $V=$ 8.5 up to $V=$ 14 mag. The stars on the red side of the MS appear to form 
a sequence parallel to the MS, that can be ascribed to Galactic disk field stars. 
Below $V\sim$ 14 mag, MS if exists, can not be isolated from the strong field star contamination. In the case of
Bochum 6, we have observed stars down to $V=$ 21 mag and contribution of field stars is more after
$V\sim$14.5 mag. It is also well known that in OB associations only O and B type stars are found and
hence probability of forming the fainter member is negligible. Therefore, we are interested only in brighter
members of the associations. Separation of the members of these associations from the field stars is a
difficult task in the absence of precise proper motion and/or radial velocity measurements for all
the stars of the region. Proper
motion measurement is available for 3 stars in Bochum 1 and 5 stars in Bochum 6 and has been used
to separate the members of these associations. The detail description about the procedure is given in
the next section. In addition to this we also adopted the photometric criteria in selection of probable
members. According to this criteria a star is considered as a member which simultaneously having
reconcilable position in colour-colour and colour-magnitude diagrams respectively. For this we defined
the blue and red envelope of the MS in colour-magnitude diagrams. In this way 12 stars in Bochum 1 and 9
 stars in Bochum 6 are separated as a probable members and they are listed in Table 4.

%
\begin{figure}
\begin{center}
\psfig{figure=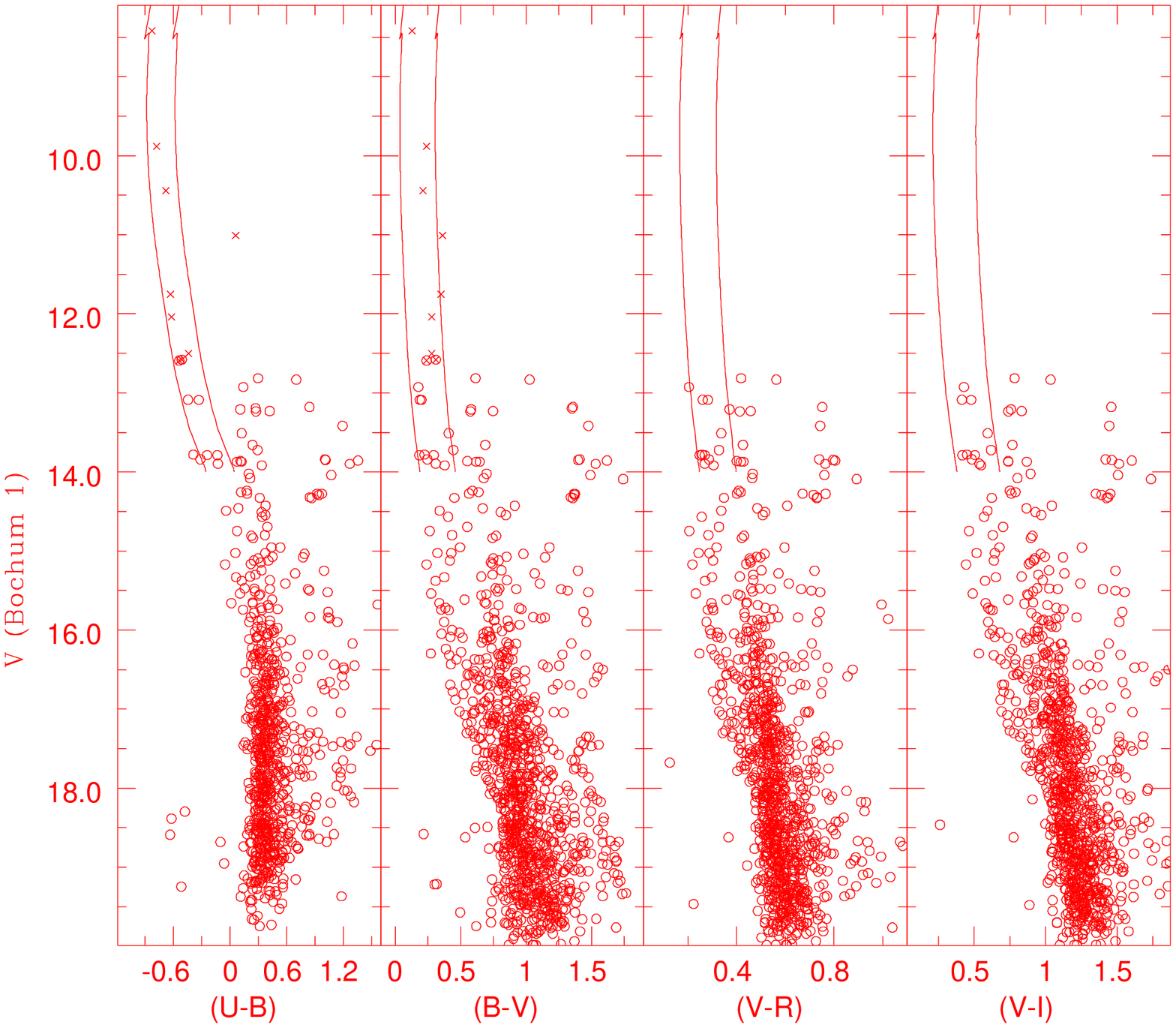,height=9cm,width=11cm}
\psfig{figure=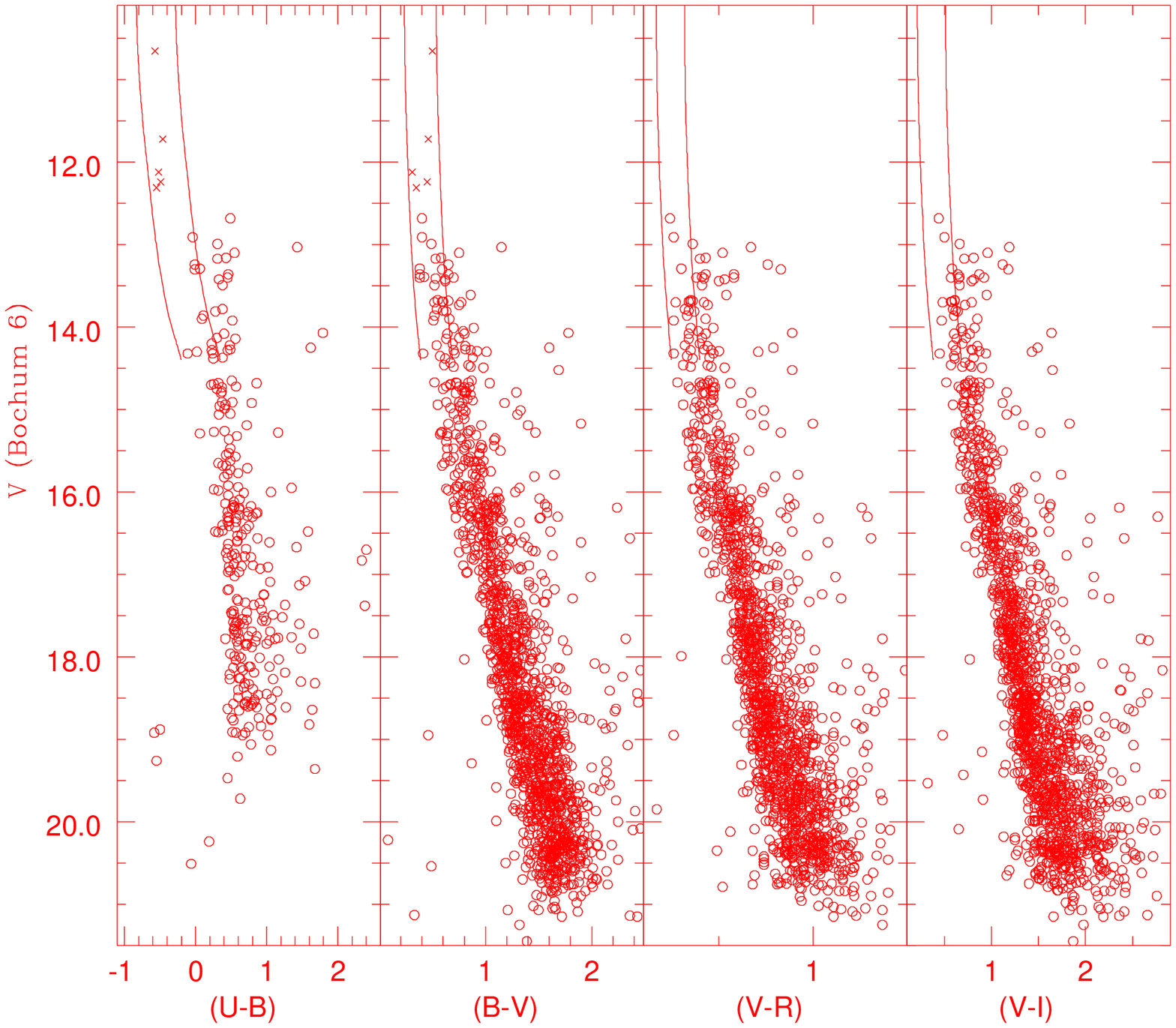,height=9cm,width=11cm}
\caption {The colour-magnitude diagrams of all the measured stars in the regions
of Bochum 1 and Bochum 6. Open circles and crosses represent the stars observed
by us and stars taken from Moffat \& Vogt (1975) for both associations respectively. 
Solid lines represent the blue and red envelope of the MS of the association.}
\label{fig3.19}
\end{center}
\end{figure}

\subsection{Evidence of clustering and membership}
 
In the Tycho-2 catalogue (H$\o$g et al. 2000), astrometric data is available
for the region of Bochum 1 and Bochum 6. The accuracy (2 mas/yr) of Tycho-2 astrometric
data is of the same order as of the Hipparcos data (Makarov et al. 2000). Because of the
large number of stars down to $V=$ 12.5 mag, the Tycho-2 catalogue may provide 
an opportunity for the separation of stars having similar motions as has been done 
earlier by Patat et al. (2001) for Bochum 9, 10 and 11 and by Carraro (2002) for NGC 133.

In the $30^{\prime} \times 30^{\prime}$ area centered on Bochum 1 and Bochum 6, Tycho-2 has 
provided the proper motion data for 30 and 50 bright ($V<$ 13 mag) stars respectively. We 
present the vector point diagrams of these stars in Fig 4. The filled circles 
denote the probable members given by Moffat \& Vogt (1975) and the horizontal and vertical 
bars represent the uncertainties in the proper motion components. A close
inspection of vector point diagram of Bochum 1 indicates that a group (including 3 stars given by 
Moffat \& Vogt (1975)) of stars are having
 approximately same proper motion component. Some stars are isolated and obviously are non-member 
of that group. In the case of Bochum 6, in spite of having large 
uncertainties in the proper motion component there seems to be a group of stars including those 5 
stars which are observed by Moffat \& Vogt (1975) having same motion 
within the error. In nutshell, we can say that there are indications that a small number of stars 
are clustered along the line of site of both Bochum 1 and Bochum 6. The average value of proper 
motion component of the group is $\mu_{\delta}$ = $-0.33\pm$1.15 and $\mu_{\alpha}$cos($\delta$)
=0.67$\pm$0.67 for Bochum 1 and $\mu_{\delta}$ = $2.68\pm$1.53 and $\mu_{\alpha}$cos($\delta$) 
=$-5.02\pm$2.24 for Bochum 6. In this way we selected 3 stars in Bochum 1 and 5 stars in Bochum 6 
having nearly same proper motion component and list them in Table 4. These stars are also 
selected as probable members using photometric method (see Sec 3.1). Finally, 15 and 14 stars in 
Bochum 1 and Bochum 6 respectively are selected as probable members using photometric as well as 
kinematical data and will be used in the further study. 

\begin{table}
\caption{Photometry and proper motion data of likely member stars in the field of Bochum 1 and 
Bochum 6. In the first column the ID number 44, 46, 47, 51, 53, 54, 48 and 45 in Bochum 1 and 499, 
501, 496, 503 and 498 in Bochum 6 represent the stars taken from Moffat \& Vogt (1975). The ID 
number starting from 1 in both associations are observed by us.}
\vspace{0.5cm}
\footnotesize
\centering
\begin{tabular}{cccccccccc}
\hline
\hline
&&&&&&\\
ID &TYC No&$(U-B)$&$(B-V)$&$(V-R)$&$(V-I)$&$V$&$\mu_{\alpha}Cos({\delta})$&$\mu_{\delta}$&$E(B-V)$\\ 
&&(mag)&(mag)&(mag)&(mag)&(mag)&(mas/yr)&(mas/yr)&(mag)\\ \hline
&&&&&&&&&\\
&&&&Bochum 1&&&&&\\
  44&133601468&  -0.83&  0.13&     *&     *&  8.42&-0.1&-0.1&0.45\\
  46&133601813&  -0.78&  0.24&     *&     *&  9.88&1.1&-1.9&0.56\\
  47&         &  -0.68&  0.21&     *&     *& 10.44&&&0.49\\ 
  51&133601173&  -0.63&  0.35&     *&     *& 11.75&1.0&1.0&0.65\\
  53&         &  -0.62&  0.28&     *&     *& 12.04&&&0.56\\
  54&         &  -0.44&  0.28&     *&     *& 12.50&&&0.52\\
  48&         &  -0.51&  0.31&     *&     *& 12.58&&&0.57\\
  45&         &  -0.54&  0.24&     *&     *& 12.59&&&0.48\\
   1&         &  -0.44&  0.19&  0.26&  0.41& 13.09&&&0.39\\
   2&         &  -0.33&  0.20&  0.28&  0.48& 13.09&&&0.36\\
   3&         &  -0.39&  0.22&  0.26&  0.45& 13.78&&&0.42\\
   4&         &  -0.24&  0.29&  0.27&  0.51& 13.79&&&0.49\\
   5&         &  -0.14&  0.18&  0.25&  0.42& 13.79&&&0.27\\
   6&         &  -0.32&  0.25&  0.29&  0.48& 13.84&&&0.41\\
   7&         &  -0.13&  0.31&  0.27&  0.54& 13.89&&&0.45\\
&&&&&&&\\
&&&&Bochum 6&\\
 499&598701023&  -0.57&  0.50&     *&     *& 10.65&-2.0& 3.7&0.82\\
 501&598700524&  -0.46&  0.46&     *&     *& 11.72&-6.6& 0.5&0.74\\
 496&598700891&  -0.52&  0.31&     *&     *& 12.12&-7.6& 1.9&0.57\\
 503&598700657&  -0.49&  0.45&     *&     *& 12.24&-5.2& 2.9&0.73\\
 498&598701067&  -0.55&  0.35&     *&     *& 12.31&-3.7& 4.4&0.63\\
   1&         &  -0.04&  0.40&  0.26&  0.50& 12.91&&&0.54\\
   2&         &   0.01&  0.65&  0.76&  1.12& 13.24&&&0.81\\
   3&         &   0.06&  0.38&  0.30&  0.58& 13.29&&&0.47\\
   4&         &   0.11&  0.52&  0.33&  0.60& 13.86&&&0.61\\
   5&         &   0.09&  0.66&  0.43&  0.85& 13.90&&&0.80\\
   6&         &   0.23&  0.69&  0.40&  0.69& 14.20&&&0.78\\
   7&         &   0.23&  0.70&  0.35&  0.62& 14.28&&&0.79\\
   8&         &  -0.11&  0.71&  0.47&  0.94& 14.32&&&0.95\\
   9&         &   0.25&  0.62&  0.36&  0.61& 14.38&&&0.71\\
\hline
\end{tabular}
\end{table}

\subsection {Colour-Colour diagrams}
 
\noindent To determine the reddening in the direction of the region Bochum 1 and Bochum 6,
 we plotted $(U-B)$ versus $(B-V)$ diagram of probable members in Fig \ref{fig3.21}.
The reddening seems to be non-uniform in the line of site of both associations. Using the ZAMS given
by Schmidt-Kaler (1982) we found that the $E(B-V)$ value ranges from 0.40 to 0.60 mag in the direction
of Bochum 1 and from 0.55 to 0.80 mag in the direction of Bochum 6. The cause of
the variability in $E(B-V)$ in these associations needs to be investigated for
understanding the star formation processes. As the stars are of early type, Q
- method is used to determine the individual $E(B-V)$ values of the probable
members and the same has been applied in generating the intrinsic
CM diagrams discussed below. The mean values of the $E(B-V)$ are
given in Table 5. They are in good agreement with the values given in the literature (see Table 1).
 
\begin{figure}
\begin{center}
\hbox{
\hspace{-1.0cm}\psfig{figure=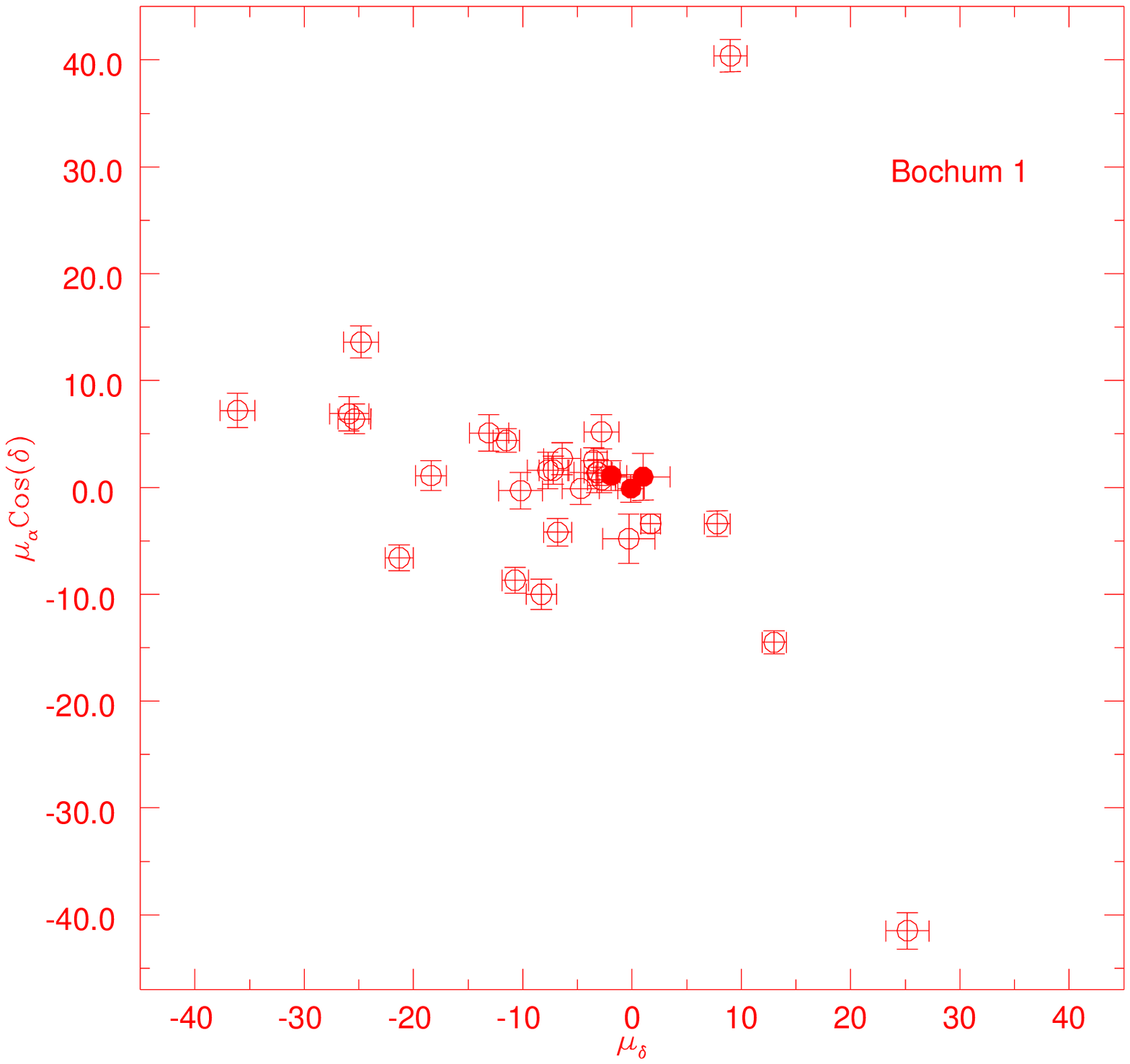,height=8cm,width=8cm}
\hspace{-1.0cm}\psfig{figure=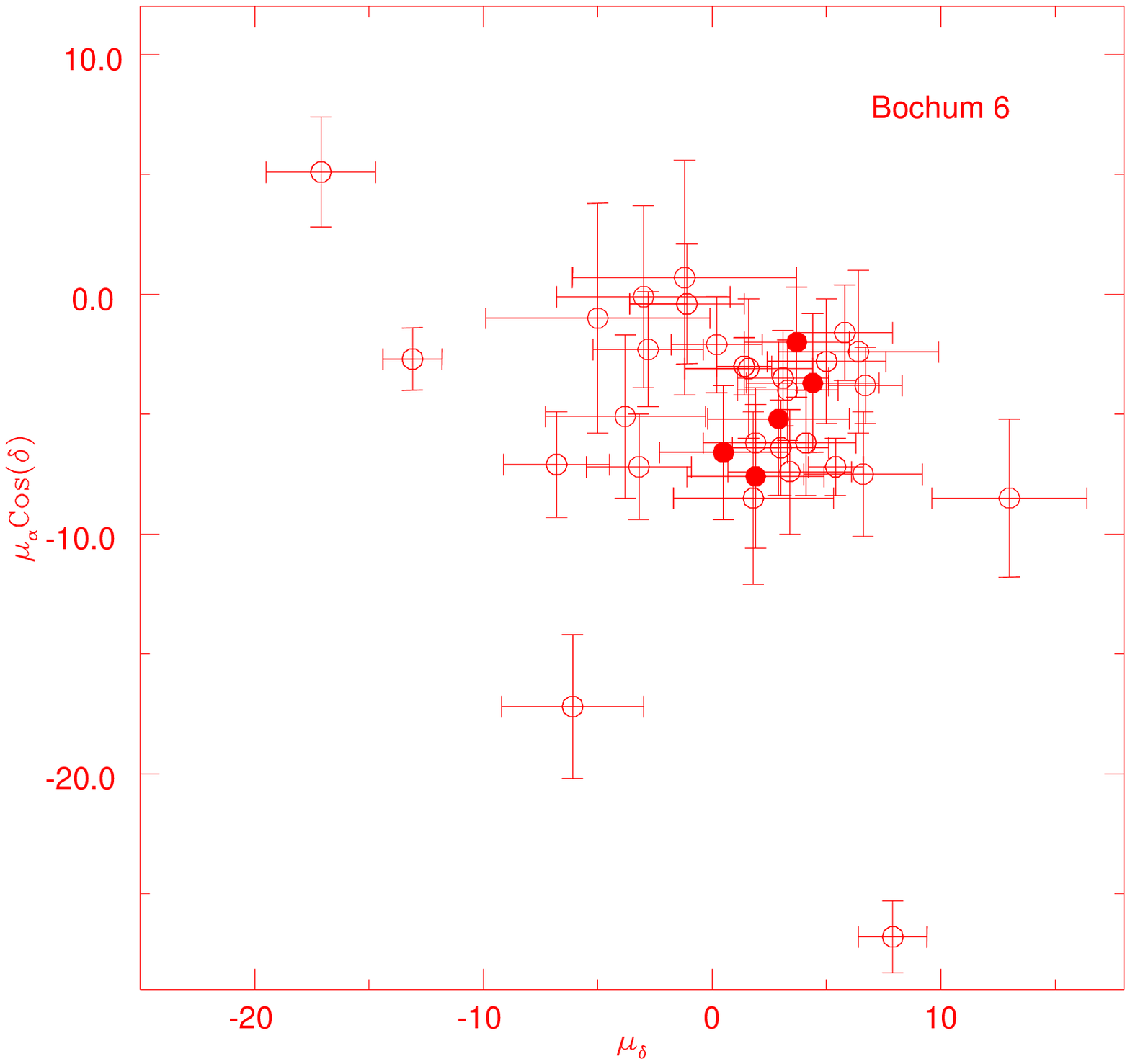,height=8cm,width=8cm}
}
\caption {Vector point diagrams for Tycho-2 stars in the filed of Bochum 1
and Bochum 6. Filled circles indicate the member of Bochum 1 and Bochum 6
according to Moffat \& Vogt (1975). The lengths of horizontal and vertical
bars indicate the uncertainties in the proper motion in $mas/yr$.}
\label{fig3.20}
\end{center}
\end{figure}
 
\subsection{Distance and Age}
 
\begin{figure}
\begin{center}
\hbox{
\hspace{-1.0cm}\psfig{figure=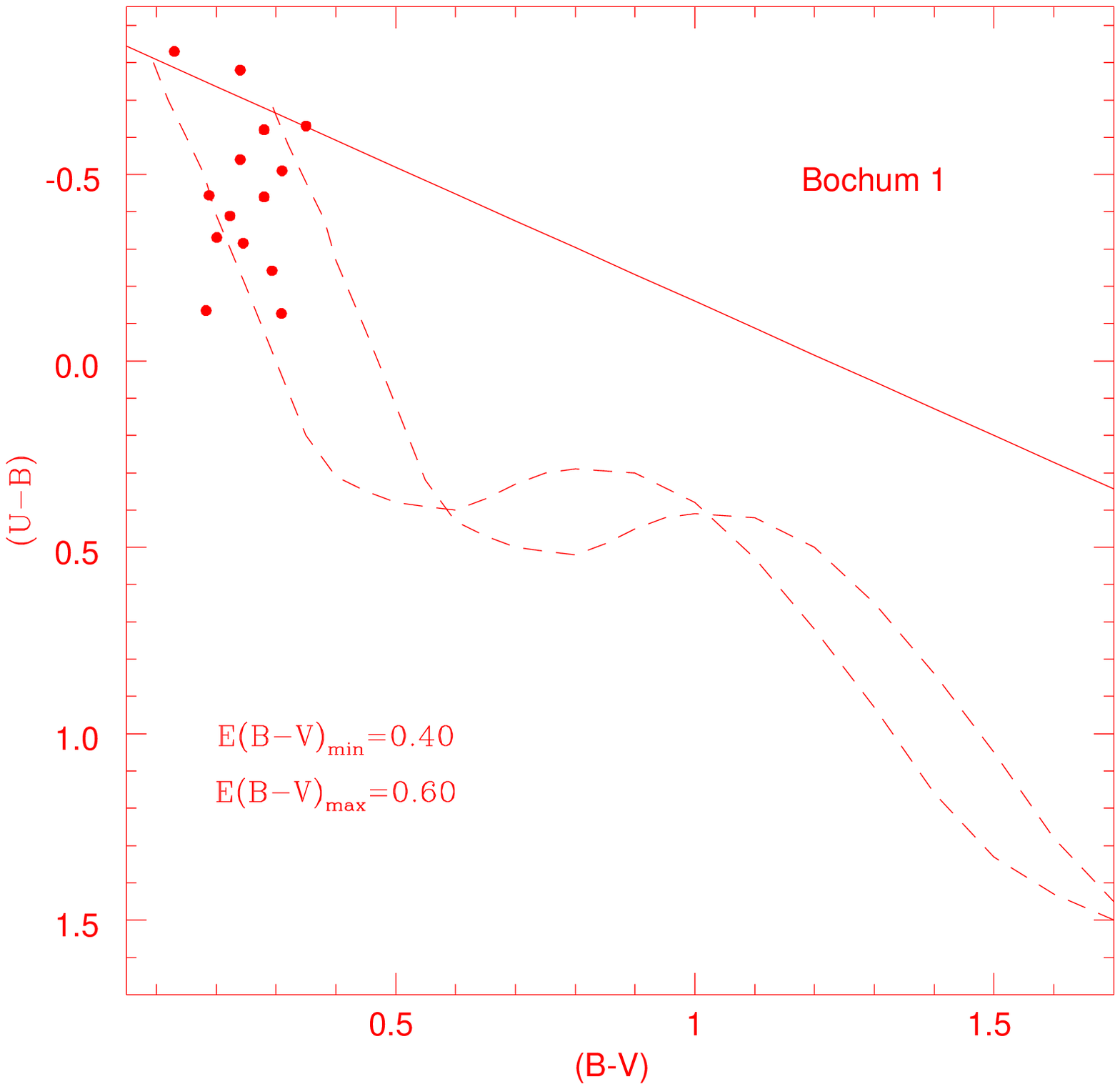,height=8cm,width=8cm}
\hspace{-1.0cm}\psfig{figure=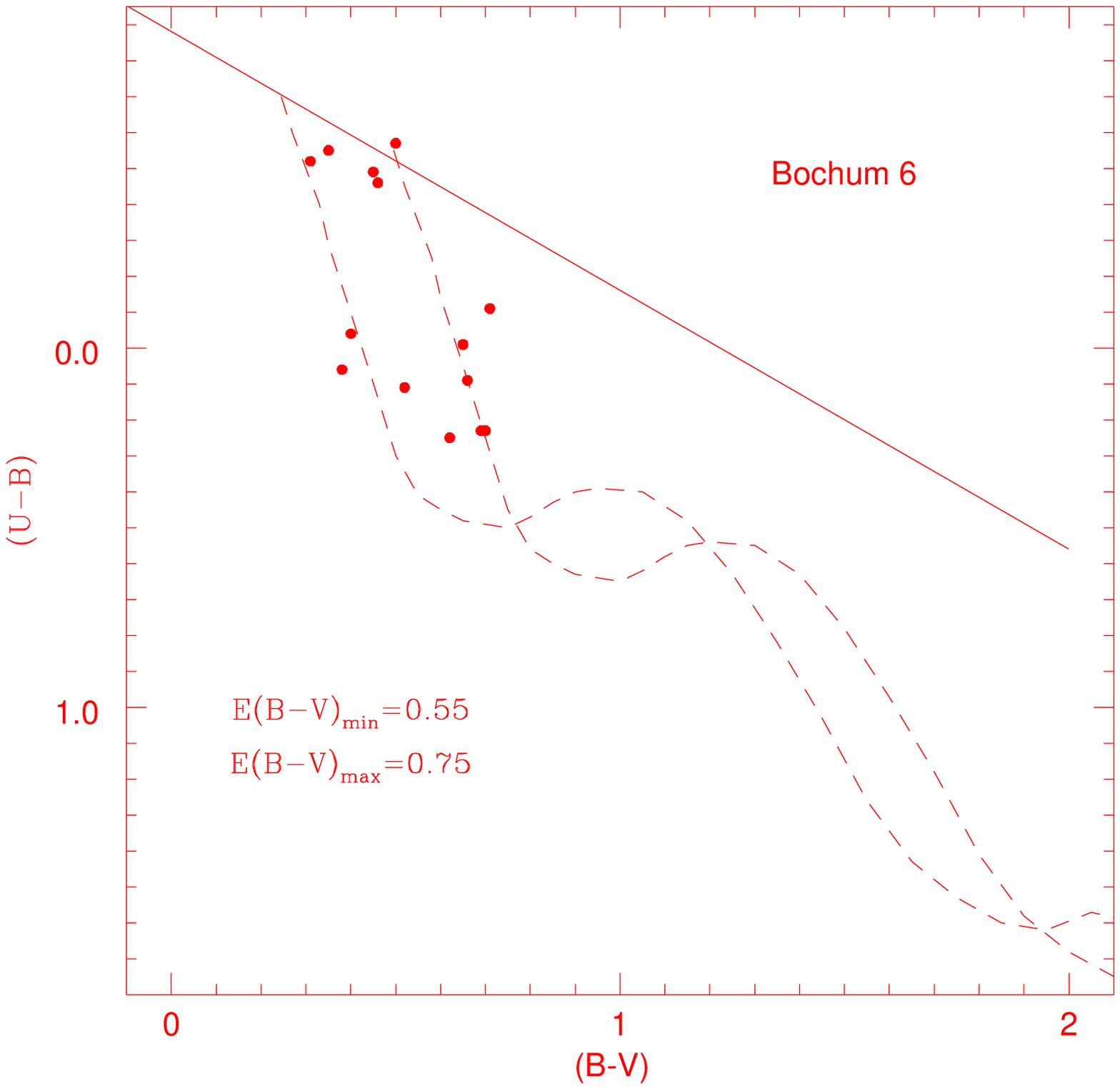,height=8cm,width=8cm}
}
\caption {The $(U-B)$ versus $(B-V)$ diagrams of the stars in the region
of Bochum 1 and Bochum 6. The continuous straight line and dotted curve
represents the reddening vector and empirical Schmidt-Kaler (1982) ZAMS.}
\label{fig3.21}
\end{center}
\end{figure}

\noindent In Fig 6, we plot the intrinsic CM diagrams in the plane of $V_{0}$ vs 
$(U-B)_{0}$ and $V_{0}$ vs $(B-V)_{0}$ for Bochum 1 and Bochum 6 to estimate their distances and ages.
The probable members listed in Table 4, have been used in this analysis. All the stars have been 
corrected for the individual reddening using $E(B-V)$ values given in Table 4 and following the 
relation $E(U-B)=0.72\times E(B-V)$ and $A_v=3.1\times E(B-V)$. In Fig, we
\begin{figure}
\begin{center}
\hbox{
\psfig{figure=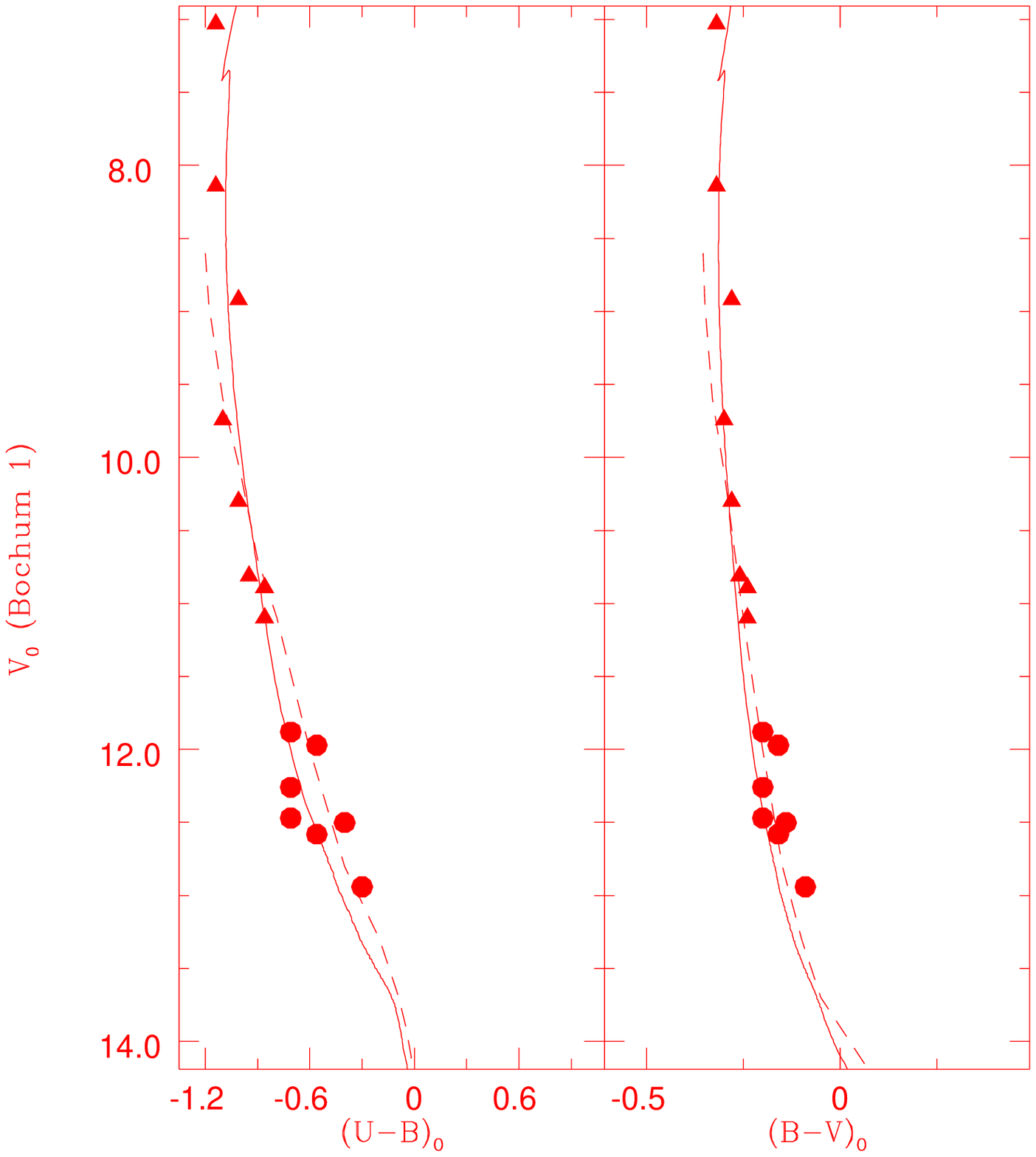,height=9cm,width=8.5cm}
\hspace{-2.0cm} \psfig{figure=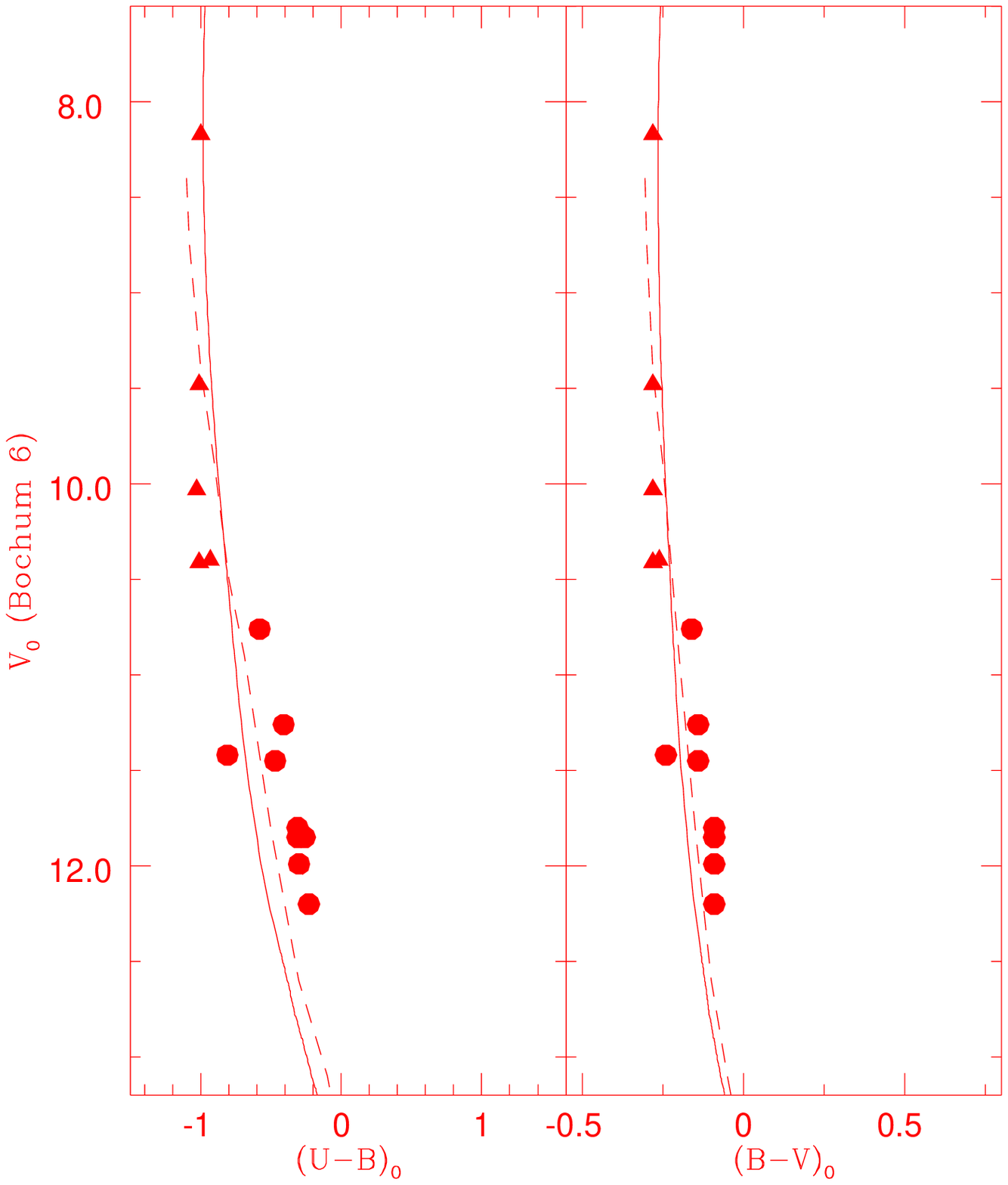,height=9cm,width=8.5cm}
}
\caption {The intrinsic CM diagrams of the probable members of associations  Bochum 1 and
Bochum 6. Filled circles and filled triangles represent probable photometric and kinematic members
. Solid lines represent the Solar metallicity isochrones given by Schaller et al. (1992) for log(age) = 7.00
while dotted lines represent the ZAMS given by Schmidt-Kaler (1982).}
\label{fig3.22}
\end{center}
\end{figure}
superimposed the ZAMS with dotted line given
by Schmidt-Kaler (1982). The fitting of ZAMS to the member stars is satisfactory and gives
the true distance modulus $(m-M)_{0}=$ 12.2$\pm0.2$ and 12.0$\pm0.2$ mag for Bochum 1 and 
Bochum 6 respectively. The corresponding distances are 2.8$\pm$0.4 and 2.5$\pm$0.4 Kpc respectively. 
The estimated distances for both the associations should 
be considered reliable because they have been derived by fitting the ZAMS over a wide range 
of MS. This is possible due to CCD photometry of stars fainter than Moffat \& Vogt (1975). 
The derived distance values for both Bochum 1 and 
Bochum 6 are lower than the corresponding values given by Moffat \& Vogt (1975)
while the value (2.5 Kpc) given by Dias et al. (2002) for Bochum 1 is in agreement.
 
 To derive the ages for Bochum 1 and Bochum 6, we have plotted the isochrones of log(age) = 7.0
taken from Schaller et al. (1992) for Solar metallicity in the Fig \ref{fig3.22}. By fitting the isochrones we have
estimated 10$\pm$5 Myr age for Bochum 1 and Bochum 6 which is in agreement with 10 Myr given 
by Moffat \& Vogt (1975).
 
\begin{table}[ht]
\caption{Derived fundamental parameters of Bochum 1 and Bochum 6.}
\label{table3.7}
\vspace{0.5cm}
\footnotesize
\begin{center}
\begin{tabular}{|c|ccc|}
\hline
\hline
Name &Distance&$E(B-V)$&log(age)\\
&(Kpc)&(mag)&\\ \hline
Bochum 1&2.8$\pm$0.4&0.47$\pm$0.10&7.0\\
Bochum 6&2.5$\pm$0.4&0.71$\pm$0.13&7.0\\
\hline
\end{tabular}
\end{center}
\end{table}

\subsection{Stellar evolutionary aspects and star formation history}
Using the parameters derived in the present study and stellar evolutionary tracks given by Schaller 
et al. (1992) for Solar metallicity we derived the masses of probable members of Bochum 1 and Bochum 6. 
The mass range of the members are 17.0$-$3.0 $M_{\odot}$ and 14.0$-$3.0
$M_{\odot}$ respectively. From the stellar evolution theory we know that 17.0 and 14.0 $M_{\odot}$
star will leave the MS after about 10 and 15 Myr respectively. On the other hand 3.0 $M_{\odot}$
star will take about 5 Myr to reach on MS. Hence the ages of brighter members are smaller than
10 and 15 Myr in the case of Bochum 1 and Bochum 6 respectively while the corresponding ages of fainter
stars are larger than 5 Myr. We can therefore conclude that
brighter and fainter members in both associations have almost similar ages and may probably have
formed together about 10 Myr ago.

\subsection{Mass Function}

The mass function (MF) of a cluster is derived from its luminosity function (LF). For 
deriving the LF, it is essential to correct the data for field star contamination and data 
incompleteness. For separating the members, we used 
the kinematical as well as photometric data (see Sec. 3.1, 3.2). In this way all the selected 
probable members are brighter than 14.5 mag in both associations and at such brightness level 
data completeness is almost 100\%. Using these probable members, we derived the LF of Bochum 1 
and 6. 

%
The MF, which denote the relative number of stars in
unit range of mass centered on mass $M$. The MF slope has been derived from the
mass distribution $\xi$($M$). If $dN$ represents the number of stars
in a mass bin $dM$ with central mass $M$, then the value of slope $x$ is
determine from the linear relation
 
\vspace{0.2cm}
\begin{center}
~~~~~~~~~~~~~~~~log$\frac{dN}{dM}$ = $-$(1+$x$)$\times$log($M$)$+$constant
\end{center} 
\vspace{0.2cm}
\noindent using the least-squares solution. The Salpeter (1955) value for the
slope of MF is $x$ = 1.35.

\begin{figure}
\hspace{1.5cm}\psfig{file=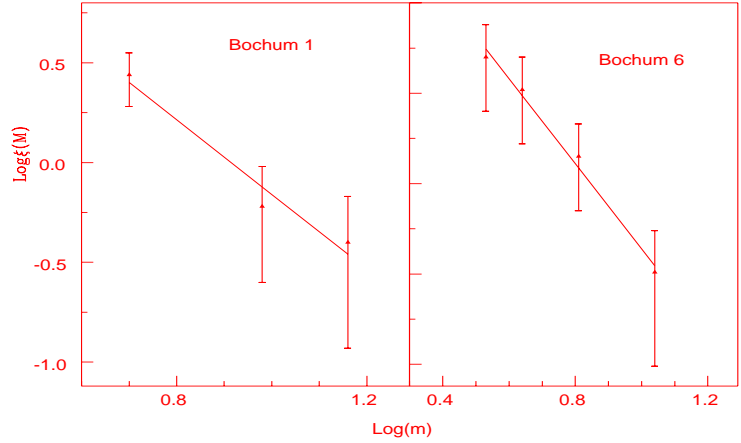 ,width=11.0cm,height=8.8cm}
\caption{The plot shows the mass functions derived using isochrones of Solar metallicity 
given by Schaller et al. (1992).}
\label{fig6}
\end{figure}

\begin{table}[h]
 \centering
\caption{The mass function slope derived from Schaller et al. (1992) isochrone of Solar metallicity.} 
\vspace{1.0cm}
\begin{tabular}{ccc}
\hline\hline
Objects&Mass range&Mass Function slope \\
&M$_{\odot}$&($x$)\\
\hline
Bochum 1&3.0 - 17.0&0.87$\pm$0.37\\
Bochum 6&3.0 - 14.0&1.35$\pm$0.16\\
\hline
\end{tabular}
\label{table4}
\end{table}

To derive the MF from LF we need theoretical evolutionary 
tracks and accurate knowledge of physical parameters of Bochum 1 and 6. Using the parameters 
derived by us and stellar evolutionary tracks of Solar metallicity given by Schaller et al. (1992) we converted the observed 
LF to the MF. Fig 7 shows the plot of MFs of Bochum 1 and 6 while Table 6 lists the estimated MF 
slope. Our derived MF slope for both Bochum 1 and 6 are in agreement with Salpeter's (1955) value, 
however the estimated errors are large due to small number of members.

Massey et al. (1995) studied 11 OB associations of the Milky Way and conclude that there is no 
statistically significant difference in slope among these objects with mean value of slope 
$x$=1.1 for stars with masses $>$ 7.0 $M_{\odot}$. They also compared the IMF study of the 
Magellanic Cloud's OB associations and suggest that there is no difference in IMF slopes. 

\section{Conclusions}

In this paper we have presented new CCD $UBVRI$ photometry for the 2460 stars in the field of 
Bochum 1 and Bochum 6. We can summarized our findings as follows.

(i) In both associations variable reddening is present 
with mean value of $E(B-V)=$0.47$\pm$0.10 and 0.71$\pm$0.13 for Bochum 1 and Bochum 6 respectively.
The corresponding distance values are 2.8$\pm$0.4 and 2.5$\pm$0.4 Kpc respectively. 
The fitting of Schaller et al. (1992) isochrones of Solar metallicity to the intrinsic 
CM diagrams indicate an age of 10$\pm$5 Myrs for both OB associations.

(ii) Present analysis indicate that brighter and fainter members 
of both associations have almost similar ages and may probably have formed 
together.

(iii) Mass function study gives MF slope of 0.87$\pm$0.37 and 1.35$\pm$0.16
for Bochum 1 and Bochum 6 respectively. They are in agreement within uncertainty with 
the Salpeter's (1955) value.

\end{document}